# Bridging the Gap between Crisis Response Operations and Systems

Khaled M. Khalil, M. Abdel-Aziz, Taymour T. Nazmy, Abdel-Badeeh M. Salem
Faculty of Computer and Information Science Ain shams University Cairo, Egypt
kmkmohamed@gmail.com, mhaziz67@gmail.com, ntaymoor@yahoo.com,
absalem@asunet.shams.edu.eg

**Abstract.** There exist huge problems in the current practice of crisis response operations. Response problems are projected as a combination of failure in communication, failure in technology, failure in methodology, failure of management, and finally failure of observation. In this paper we compare eight crisis response systems namely: DrillSim [2, 13], DEFACTO [12, 17], ALADDIN [1, 6], RoboCup Rescue [11, 15], FireGrid [3, 8, 18], WIPER [16], D-AESOP [4], and PLAN C [14]. Comparison results will disclose the cause of failure of current crisis response operations (the response gap). Based on comparison results; we provide recommendations for bridging this gap between response operations and systems.

## 1  Introduction

The crisis response domain is characterized as a virtual environment of required distributed control, huge amount of data, uncertainty, ambiguity, multiple stakeholders with different objectives, and limited resources which continually vary [1]. In consequence of mentioned domain characteristics; crisis response systems require a multi-disciplinary system design approach.

One of the crisis response systems design approaches is to mimic a crisis by conducting crisis drills over a sample region; incorporating information technologies in the process of response during the drill. Drills are expensive and scripted to given crisis situations. Also, large scale testing solutions are close to impossible to test via drills [2].

Another approach is to use simulation and modeling tools. Simulation and modeling tools allow creating what-if scenarios dynamically and determining the ability of the response to adapt to the changing crisis requirements. Actually, simulation and modeling approach has an extra benefit that reliable simulation model can be used for real-time support operations enhancing situational awareness and decision support [9]. Simulation and modeling systems for crisis response consist of a set of integrated tools which will differ based on the application they are designed for (Fig 1). Based on the definition of

Integrated Emergency Response Framework (iERF), simulation and modeling tools include six types of tools. Planning tools are used for determination of impact of a crisis event, and/or aiding development of the response action plans and strategies. Vulnerability analysis tools are used for evaluation and assessment of response preparedness plans. Identification and detection tools are used for determining the possibility of the occurrence of crisis event Training tools are used for training response personnel for handling crisis events. Systems testing tools are used for testing of systems and equipments used for crisis response. Real-time response support tools are used for evaluation of the current/future impact of a crisis through real-time updates on the situation, and evaluation of alternative actions/strategies evaluations which are then used to direct the response actions on the ground.

The scope of the simulation tools can vary from national level modeling for large disaster events such as volcanic explosions, to modeling a city block for a scenario like a building explosion or fire.

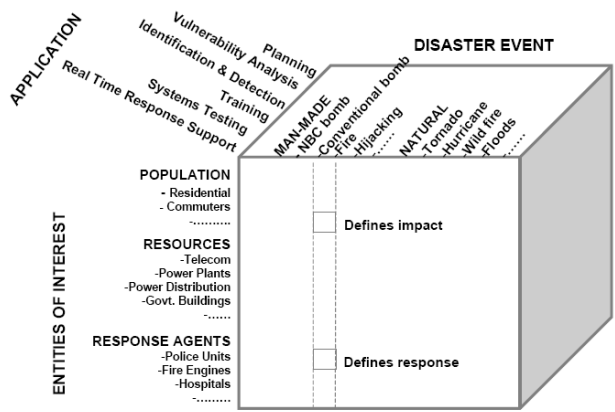

**Fig. 1.** Integrated Emergency Response Framework (iERF) proposed by NIST [9]

In what follows, we present systems comparisons of current response systems which are based on iERF and design requirements satisfaction. Then we recommend a set of actions to bridge the response gap.

## 2  Systems Comparison

We have selected eight response systems: DrillSim, DEFACTO, ALADDIN, RoboCup Rescue, FireGrid, WIPER, D-AESOP, and PLAN C for comparison to discover the operations and systems gap. At first we have to project systems on the iERF Framework axes to recognize how system response to crisis events (see table 1).

**Table 1.** Crisis Response Systems Comparison based on iERF

| System | Purpose | Crisis Event | Entities (Agents) Methodology | Main Features |
|---|---|---|---|---|
| DrillSim | Testing IT solutions | Fire Evacuation | - Agent behaviour has been modelled as a discrete process | -Integration with other simulations<br>-Calibration of agent behaviour<br>-Dynamic planning |
| DEFACTO | Improving situation awareness in response operations | Fire Evacuation | - Agent has been modelled in proxy team formation | - 3D visualization<br>- Adjustable autonomy<br>-Conflict resolution<br>- Proxy framework |
| ALADDIN | Model decentralized systems that can bring together information from variety of heterogeneous sources in order to take informed action | Fire Evacuation | - System is composed of reactive and proactive agents.<br>- Agents can sense, act and interact in order to achieve individual and collective goals | - Sensors network<br>- Minimal agent communication<br>- adaptive on-line decision making<br>- Data fusion techniques |
| RoboCup Rescue | Large-scale simulation for urban-search and rescue | Search and rescue | - Agents collect, store, and evaluate information.<br>- Agents choose best actions fitting to the situation | - space-exploration techniques<br>- Prediction for the civilian's life time<br>- Calibration of agent teams behavior |
| FireGrid | Pursue research for developing real-time response systems using the Grid | Fire Evacuation | Command-and-control (C2) tasks. | - Self-Configuring sensors network<br>- High level plan<br>- Agent safety and security |
| WIPER | Evaluate potential plans of action using a series of GIS enabled Agent-Based simulations | Predict simple movement and traffic patterns | -Web Service and Service Oriented Architecture<br>- Multi-Agent System Design | - Anomaly detection algorithm flagging potential crisis<br>- Predict the course of events |
| D-AESOP | Model of situation awareness into the environment of BDI agent based MAS | Medical relief operations | - Extended BDI Agent Model<br>- Medical relief ontology | - Situation recognition process<br>- Event situation plan<br>- Extended BDI Model |
| PLAN C | Improve planning and response to the public health and medical consequences of a mass casualty event | Public health and medical relief operations | A large number of agents: Person, Hospital, On-Site Responder, Ambulance and Catastrophe | - Integration of medical, and public health in the model<br>- Realistic models of medical and responder units effects<br>- Integration of GIS data |

### 2.1 The Response Gap

Given the first comparison results (system purpose, agent methodology, and main feature), we need to identify the design requirements of response operations. From previous work [10], we have ten design requirements for crisis response systems. Table 2 shows systems comparison based on these systems design requirements.

**Table 2.** Systems Comparison based on systems design requirements [10]

| System | Response Effectiveness | Standard (NIST or IEEE) | Design Requirements ||||||||
|---|---|---|---|---|---|---|---|---|---|---|---|
| | | | Control | Communication | Resources | Information | Actors Action | Adaptation | Planning | Integration | Learning |
| DrillSim | | | | | | ● | | | | ● | ● |
| DEFACTO | | | ● | | | ● | | | | | |
| ALADDIN | ● | | ● | ● | ● | ● | | | ● | | ● |
| RoboCup Rescue | | | ● | | | ● | | | ● | | ● |
| FireGrid | | | ● | | | ● | | | | ● | |
| WIPER | | | ● | | ● | ● | | | ● | ● | ● |
| D-AESOP | | | ● | | ● | ● | | | ● | | ● |
| PLAN C | | | | ● | ● | ● | | | ● | | ● |

Table 2 comparison shows the gap: (i) Set of systems had failed to manage overloaded communication where others had failed to manage dynamic resources over time. (ii) None of the compared systems had dealt with adaptation of system components to environment changes or had dealt with tracking actors' actions. (iii) Current systems had focused on roughly supporting response activities with small interest on improving the effectiveness of response operations. (iv) Systems development doesn't follow any standards in spite of existing standards waiting to be adopted in response systems. (v) The recognized response gap will degrade the effectiveness or even stop response operations in case of damaged systems components.

## 2.2 Recommendations

Design requirements [10] are proposed as the core functionality required by future crisis response systems. Design requirements focus on the high availability of response system beside the adaptation of system behavior and components to envirenement changes. Failure of communication is tackled by exposing minimal communication and supporting intelligent communication devices which adapt with available spectrums. Failure of technology is tackled by integration with other components and systems. Failure of methodology is tackled by providing adaptive planning and resource management. Failure of management is tackled by distributed control and natural decision making. And finally, failure of observation is tackled by gathering and fusing of information from different data sources.

In addition, systems development should follow standards available like NIST or IEEE [9] to improve the way systems integrate and exchange information.

## 3 Conclusion and IMAMCR Project Status

The state-of-art practice of crisis response systems confronts several challenges due to the nature of crisis events. Challenges include failure of communication, technology, methodology, management, and observation. Crisis response systems development should take in consideration the design requirements of response systems to overcome the response gap and to improve response effectiveness. In addition, following systems standards will simplify the integration and exchange of information among systems.

We are working on IMAMCR (Intelligent Multi-Agent Model for Crisis Response). IMAMCR is a self-defensible and adaptable response model based on the metaphor of Artificial Immune System for pandemic flu response. IMAMCR consists of two parts; (i) Decision maker view, and (ii) the response system. The decision maker view is based on GISTool kit [7] to monitor the spread of pandemic flu through interactive map of Egypt. IMAMCR design is considered to follow the response design requirements. Thus, we have surveyed different agent architectures and frameworks to support our goal. We have selected Cougaar Agent Architecture [5]. Cougaar is a middle-ware for building agent-based applications. Cougaar supports minimal communication, different communication protocols, system components tolerance, distributed control, integration with other systems, agent learning, and agent planning.

# References

1. Adams, M. Field, E. Gelenbe, D. J. Hand: The Aladdin Project: Intelligent Agents for Disaster Management. IARP/EURON Workshop on Robotics for Risky Interventions and Environmental Surveillance, 2008.
2. Balasubramanian, Massaguer, Mehrotra, Venkatasubramanian. DrillSim: A Simulation Framework for Emergency Response Drills. Proc. of ISCRAM2006.
3. Berry, D., Usmani, A. FireGrid: Integrated Emergency Response and Fire Safety Engineering for the Future Built Environment. UK e-Science Programme All Hands Meeting, Nottingham, UK, Sept. 19-22, 2005.
4. Buford, G. Jakobson, L. Lewis, N. Parameswaran, P. Ray. D-AESOP: A Situation-Aware BDI Agent System for Disaster Situation Management. First International Workshop on Agent Technology for Disaster Management (ATDM) May 2006, Hakodate, Japan.
5. Cougaar Agent Architecture, http://www.cougaar.org/
6. Gianni, Georgios Loukas, Erol Gelenbe. A Simulation Framework for the Investigation of Adaptive Behaviors in Largely Populated Building Evacuation Scenarios. The International Workshop on "Organized Adaptation in Multi-Agent Systems, at AAMAS 2008.
7. GISToolKit, http://gistoolkit.sourceforge.net/
8. I-Rescue project, http://www.i-rescue.org/
9. Jain, Charles McLean. Modeling and Simulation for Emergency Response: Workshop Report, Standards and Tools .U.S DEPARTMENT OF COMMERCE, 2003.
10. Khalil, M. Abdel-Aziz, Taymour T. Nazmy, Abdel-Badeeh M. Salem. Multi-Agent Crisis Response systems – Design Requirements and Analysis of Current Systems. ICICIS 2009.
11. Kleiner, B. Steder, C. Dornhege, D. Höfer. RoboCupRescue - Robot League Team RescueRobots Freiburg (Germany). RoboCup (Osaka, Japan), 2005.
12. Marecki, Nathan Schurr, Milind Tambe. Agent-based Simulations for Disaster Rescue Using the DEFACTO Coordination System. Emergent Information Technologies and Enabling Policies for Counter Terrorism, 2005.
13. Massaguer, Vidhya Balasubramanian, Sharad Mehrotra, Nalini Venkatasubramanian. MultiAgent Simulation of Disaster Response.AAMAS'06 May 8–12 2006, Japan.
14. Plan C Project, http://www.nyu.edu/ccpr/laser/plancinfo.html
15. RoboCup project, http://www.robocup.org/
16. Schoenharl, Greg Madey, G´abor Szab´o, et, al. WIPER: A Multi-Agent System for Emergency Response. Proceedings of the 3rd International ISCRAM Conference (B. Van de Walle and M. Turoff, eds.), Newark, NJ (USA) 2006.
17. Schurr, Janusz Marecki, Tambe, J. P. Lewis. The Future of Disaster Response: Humans Working with Multiagent Teams using DEFACTO. AAAI Spring Symposium on "AI Technologies for Homeland Security," 2005.
18. Wickler, G., Tate, A. and Potter, S. Using the <I-N-C-A> Constraint Model as a Shared Representation of Intentions for Emergency Response. Proceedings of ATDM, 2006.